\documentclass[preprint,showpacs,titlepage,aps,prd,
tightenlines,amsmath,byrevtex,nofootinbib]{revtex4}

\usepackage{graphicx}

\def\lsim{\raise0.3ex\hbox{$\;<$\kern-0.75em\raise-1.1ex
\hbox{$\sim\;$}}}
\def\gsim{\raise0.3ex\hbox{$\;>$\kern-0.75em\raise-1.1ex
\hbox{$\sim\;$}}}

\begin{document}

\preprint{hep-ph/0601258}
\preprint{KEK-TH-1070}

\title{Resolving $\theta_{23}$ Degeneracy by \\
Accelerator and Reactor Neutrino Oscillation Experiments
}


\author{K.~Hiraide$^{1}$}
\email{hiraide@scphys.kyoto-u.ac.jp}
\author{H.~Minakata$^{2}$}
\email{minakata@phys.metro-u.ac.jp}
\author{T.~Nakaya$^{1}$}
\email{nakaya@scphys.kyoto-u.ac.jp}
\author{H.~Nunokawa$^{3}$}
\email{nunokawa@fis.puc-rio.br} 
\author{H.~Sugiyama$^{4}$}
\email{hiroaki@post.kek.jp}
\author{W.~J.~C.~Teves$^{5}$}
\email{teves@fma.if.usp.br}
\author{R.~Zukanovich Funchal$^{5}$}
\email{zukanov@if.usp.br}
\affiliation{
$^1$Department of Physics, Kyoto University, 606-8502, Kyoto, Japan \\
$^2$Department of Physics, Tokyo Metropolitan University, Hachioji, 
Tokyo 192-0397, Japan \\
$^3$Departamento de F\'{\i}sica, Pontif{\'\i}cia Universidade Cat{\'o}lica 
do Rio de Janeiro, C. P. 38071, 22452-970, Rio de Janeiro, Brazil \\
$^4$Theory Group, KEK, Tsukuba, Ibaraki 305-0801, Japan\\
$^5$Instituto de F\'{\i}sica, Universidade de S\~ao Paulo, 
 C.\ P.\ 66.318, 05315-970 S\~ao Paulo, Brazil
}

\date{May 25, 2006}

\vglue 1.4cm

\begin{abstract}

If the lepton mixing angle $\theta_{23}$ is not maximal, there arises a 
problem of ambiguity in determining $\theta_{23}$ due to the 
existence of two degenerate solutions, 
one in the first and the other in the second octant. 
We discuss an experimental strategy for resolving the $\theta_{23}$ 
octant degeneracy by combining reactor measurement of $\theta_{13}$ 
with accelerator $\nu_{\mu}$ disappearance and 
$\nu_{e}$ appearance experiments.
The robustness of the $\theta_{23}$ degeneracy and the 
difficulty in lifting it only by accelerator experiments with conventional 
$\nu_{\mu}$ (and $\bar{\nu}_{\mu}$) beam 
are demonstrated by analytical and numerical treatments. 
Our method offers a way to overcome the difficulty and can resolve 
the degeneracy between solutions $\sin^2 \theta_{23} = 0.4$ and 
$\sin^2 \theta_{23} = 0.6$ if $\sin^2 2 \theta_{13} \gsim 0.05$ at 95\% CL 
by assuming the T2K phase II experiment and a reactor measurement 
with an exposure of 10 GW$\cdot$kt$\cdot$yr.
The dependence of the resolving power of the octant degeneracy 
on the systematic errors of reactor experiments is also examined.

\end{abstract}




\maketitle


\section{introduction}\label{introduction}

The atmospheric neutrino observation by Super-Kamiokande (SK),  
which discovered neutrino oscillation, established that the mixing angle 
$\theta_{23}$ is large, which may be even close to the maximum \cite{SKatm}. 
Then, it became clear that the solar mixing angle $\theta_{12}$ 
is also large, though not maximal, 
as indicated by accumulation of the solar and the reactor data 
\cite{solar,KamLAND}. 
Curiously enough, the third angle $\theta_{13}$ is known to be 
small \cite{CHOOZ,K2K_app}.
Comprehending the structure of lepton mixing described by 
the Maki-Nakagawa-Sakata (MNS) matrix \cite{MNS} 
which is composed of a nearly maximal, a large, and a small angle 
should reveal the key to understand the underlying physics of 
lepton flavor mixing.

Toward the explanation of the coexistence of nearly maximal and small 
mixing angles some symmetries  \cite{symmetry} which are motivated by 
more phenomenological $\mu \leftrightarrow \tau$ exchange 
symmetries \cite{mu-tau} are discussed.
Interestingly, most of them share an attractive feature that 
$\theta_{23}=\pi/4$ and $\theta_{13}=0$ 
in the symmetry limit, suggesting a possible common origin of 
the two small quantities in the lepton flavor mixing, 
$\theta_{13}$ and $D_{23} \equiv \frac{1}{2} - \sin^2 \theta_{23}$, 
the parameter which measures deviation from the maximal mixing.  
The quark-lepton complementarity \cite{QLC}, if extends to the 2-3 sector, 
might be relevant for maximal or nearly maximal $\theta_{23}$. 
If these small quantities take non-vanishing values it should give 
us a further hint for deciding if the symmetries are the right explanation 
for the two small quantities, and if so, for identifying the correct symmetry.

In this paper, we describe an experimental strategy for determining 
$\theta_{23}$ following \cite{MSYIS}, where it was proposed that 
the $\theta_{23}$ octant degeneracy can be lifted by 
combining reactor measurement of $\theta_{13}$ with 
accelerator appearance and disappearance measurement 
of certain combinations of $\theta_{23}$ and $\theta_{13}$.
See Ref.~\cite{octant} for an earlier suggestion of this 
possibility.  
We thoroughly examine this method by taking a concrete 
(and probably the best thinkable) setting for accelerator experiments, 
phase II of the T2K experiment \cite{JPARC}, 
with Hyper-Kamiokande as a detector \cite{nakamura} 
and a 4 MW neutrino beam from the J-PARC facility. 
For knowledges of reactor measurement, we are benefited by 
the outcome of the world-wide effort \cite{reactor_white}. 
We carry out a detailed quantitative analysis to 
obtain the region of $\theta_{13}$ and $\theta_{23}$ in which 
the $\theta_{23}$ octant degeneracy can be resolved by our method. 
For a related analysis, see \cite{shaevitz}.

The $\theta_{23}$ octant degeneracy is a part of a larger structure 
called the parameter degeneracy. 
It may be understood as the intrinsic degeneracy \cite{Burguet-C} 
duplicated by the unknown sign of $\Delta m^2_{31}$ \cite{MNjhep01} 
and $\theta_{23}$ octant ambiguity \cite{octant}, which lead to the 
total eight-fold degeneracy \cite{BMW1}. 
(See \cite{MNP2} for exposition of analytic structure of the degeneracies.)
Then, one has to address a problem; 
Is it possible to resolve only the $\theta_{23}$ degeneracy, 
leaving the other two types of degeneracies untouched?
We will answer in the positive this question in the 
context of conventional $\nu_{\mu}$ beam experiments. 
In short, in modest baseline distances for which perturbative treatment 
of the matter effect is valid, the $\theta_{23}$ degeneracy is 
decoupled from the other degeneracies and can be resolved 
independently in the presence of them.

The current bounds on these small quantities in the lepton flavor 
mixing are rather mild, 
\begin{eqnarray}
-0.14 \leq 
D_{23} \equiv \frac{1}{2} - \sin^2 \theta_{23}
\leq 0.14, 
\label{cbound23}
\end{eqnarray} 
at 90\% CL \cite{SKatm}, which is nothing but a translation of the bound 
$\sin^2 2\theta_{23} \geq 0.92$. 
At present, there is neither indication of deviation from the maximal 
$\theta_{23}$, nor preference of the particular octant 
(apart from very slight preference of the 2nd octant) 
in the analysis by the SK group with their current data set \cite{kajita}.
On the other hand, the bound on $\theta_{13}$ is much stronger 
if one uses the same variable, 
\begin{eqnarray}
\sin^2 \theta_{13}
\leq 0.022 ~(0.047) 
\label{cbound13}
\end{eqnarray} 
at 90\% (3$\sigma$) CL for 1 degree of freedom, 
as obtained by the global analysis \cite{global} 
with use of all the data including the Chooz, 
the atmospheric, and the K2K \cite{K2K} one.   
As discussed in \cite{MSS04} (see also \cite{munich}), 
one of the major difficulties for accurate measurement of 
$\theta_{23}$ with accelerator neutrinos is the 
$\theta_{23}$ degeneracy. 
Hence, we expect that our method will help to improve the situation.

In Sec.~\ref{how}, we explain how the $\theta_{23}$ octant degeneracy 
can be resolved by our method.   
In Sec.~\ref{robust}, we discuss how robust is the $\theta_{23}$ 
octant degeneracy by indicating the difficulties in resolving it 
only by accelerator measurement. 
In Sec.~\ref{analysis_method}, we fully explain the statistical 
procedure of our analysis. 
In Sec.~\ref{analysis_results}, we present the results of our analysis. 
In Sec.~\ref{conclusion}, we give concluding remarks. 
In Appendix~\ref{number}, we describe some details of how 
the referred numbers of events are computed in our paper.

\section{The method and what is new?}\label{how}

\begin{figure}[htbp]
\vglue -1.6cm
\begin{center}
\includegraphics[width=0.96\textwidth]{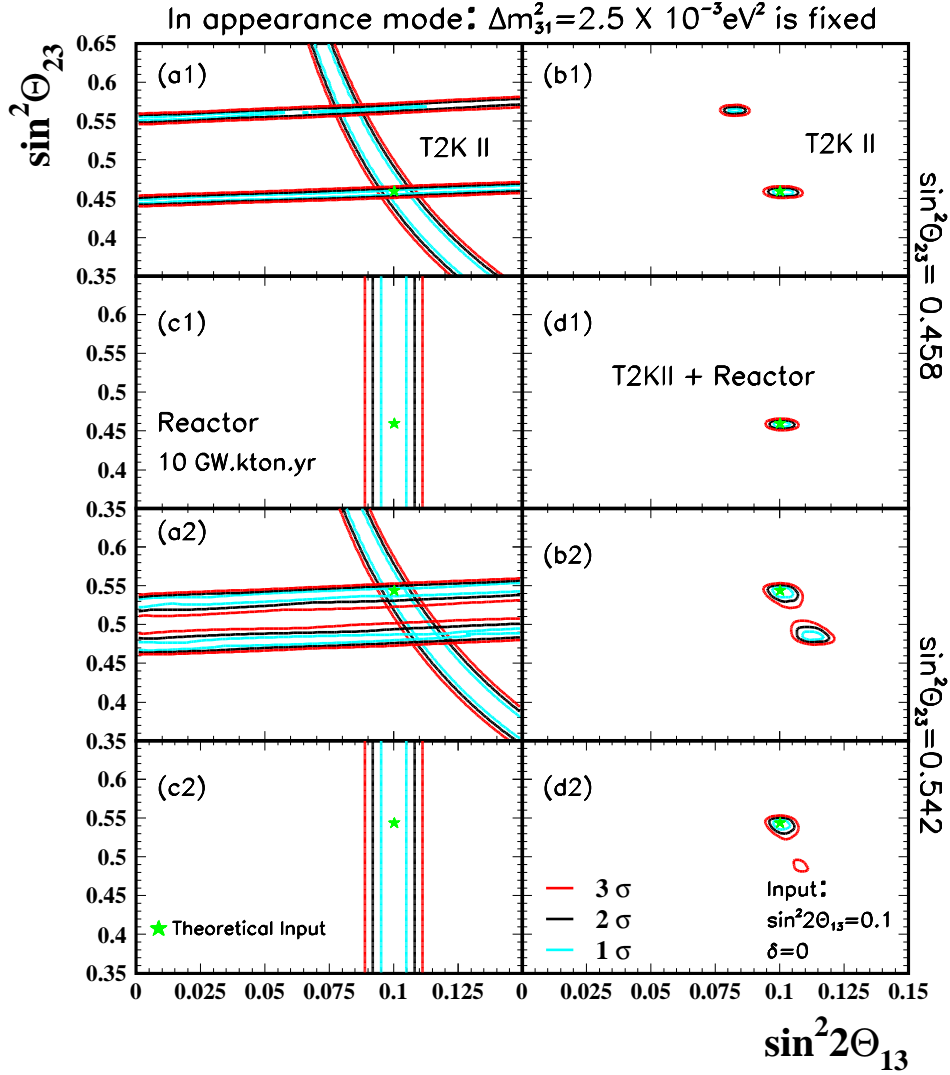}
\end{center}
\vglue -1.6cm
\caption{
The upper (lower) four panels describe the process of how the $\theta_{23}$ 
octant degeneracy can be resolved for the case where the true value 
of $\sin^2\theta_{23}$ = 0.458 (0.542), corresponding to 
$\sin^2{2\theta_{23}}=0.993$. The other input mixing
parameters are given as 
$\Delta m^2_{31} = 2.5\times 10^{-3}$ eV$^2$, 
$\sin^22 \theta_{13}$ = 0.1 and $\delta =0$, 
$\Delta m^2_{21} = 8.0\times 10^{-5}$ eV$^2$, 
$\sin^2 \theta_{12}$ = 0.31 (the input values of 
$\sin^22 \theta_{13}$ and $\sin^2\theta_{23}$ 
are indicated by the symbol of star in the plot).
(a) The regions enclosed by the solid and the dashed curves are 
allowed regions only by the results of appearance and disappearance 
accelerator measurement, respectively. 
(b) The regions that remain allowed when results of appearance 
and disappearance measurement are combined. 
(c) The regions  allowed by reactor measurement.
(d) The regions allowed after combining the results of appearance and
disappearance accelerator experiments with the reactor measurement.
The exposures for accelerator are assumed to be 2 (6) years of neutrino 
(anti-neutrino) 
running with 4 MW beam power with Hyper-Kamionande whose fiducial 
volume is 0.54 Mt, whereas for the reactor we assume an exposure of 
10 GW$\cdot$kt$\cdot$yr. 
The case of optimistic systematic error is taken. 
(See Secs.~\ref{analysis_method} and Appendix~\ref{number} for details.) 
}
\label{23fig1}
\end{figure}

For completeness of the presentation, we start by reviewing the 
method for solving $\theta_{23}$ octant degeneracy described in 
\cite{MSYIS}. 
At the end of this section, we will try to elucidate the 
difference between this work and the reference \cite{MSYIS}. 
We invite the readers to look at Fig.~\ref{23fig1}, 
and first focus on the upper four panels, the case where the input 
value of $\theta_{23}$ is in the first octant.
(The upper four and the lower four panels of Fig.~\ref{23fig1} are 
for input values of $s^2_{23}=0.458$ and 0.542, respectively.) 
Fig.~\ref{23fig1}a describes the constraints imposed by 
each accelerator experiment, $\nu_{\mu}$ (and $\bar{\nu}_{\mu}$)
disappearance and 
$\nu_{e}$ (and $\bar{\nu}_{e}$) appearance measurement. 
Although these contours come from our full analysis whose details will be 
explained in Secs. \ref{analysis_method} and \ref{analysis_results}, 
the main features of Fig.~\ref{23fig1} can be understood 
by the vacuum oscillation approximation, and essentially it is 
all that we need.  
%

The $\nu_{\mu}$ disappearance and $\nu_{e}$ appearance probabilities in 
one $\Delta m^2$ dominance approximation \cite{one-mass}, 
which may be justified by 
$\Delta m^2_{21} / \Delta m^2_{31} \simeq 1/30 \ll 1 $, are given by 
\begin{eqnarray}
P(\nu_{\mu} \rightarrow \nu_{\mu}) & = & 1 - 
\sin^2 2\theta_{23}\sin^2\Bigl(\frac{\Delta m^2_{31} L}{4E} \Bigr), 
\label{Pvac_mumu}
\\
P(\nu_{\mu}  \rightarrow \nu_{\rm e}) &=&
s^2_{23} \sin^2{2\theta_{13}} 
\sin^2 \left(\frac{\Delta m^2_{31} L}{4 E}\right), 
\label{Pvac_mue}
\end{eqnarray}
where $E$ denotes the neutrino energy and $L$ is the baseline distance. 
We use the standard notation of the MNS matrix including the 
symbol $s_{ij}$ for $\sin \theta_{ij}$ \cite{PDG}. 
$\Delta m^2_{ji}$ is defined as 
$\Delta m^2_{ji} \equiv m^2_{j} - m^2_{i}$ by using the neutrino masses 
$m_{i}$ ($i=$ 1-3).
In this approximation 
$P(\nu_{\alpha}  \rightarrow \nu_{\beta}) = 
P(\bar{\nu}_{\alpha} \rightarrow \bar{\nu}_{\beta})$. 
Accelerator disappearance measurement is expected to determine 
both $\Delta m^2_{31}$ and $\sin^2{2\theta_{23}}$
with high accuracies. 
It is obvious that, if $\theta_{23} \neq \pi/4$, one has two-fold
solutions of $\theta_{23}$ for a given value of
$\sin^2{2\theta_{23}}$; $s^2_{23} = \frac{1}{2} \left[ 1 \pm \sqrt{1-
\sin^2{2\theta_{23}}} \right]$.  This is a simple illustration of how
$\theta_{23}$ octant degeneracy arises.  The almost horizontal two
lines in Fig.~\ref{23fig1}a are nothing but these two solutions of
$s^2_{23}$ for a given value of $\sin^2{2\theta_{23}}=0.993$.

On the other hand, accelerator appearance measurement 
determines a particular combination of two angles, 
$s^2_{23} \sin^2{2\theta_{13}}$, as seen in (\ref{Pvac_mue}) 
for a given value of $\Delta m^2_{31}$. The latter quantity is 
expected to be well determined by disappearance measurement. 
The curved strip in Fig.~\ref{23fig1}a, which takes the approximate 
form $s^2_{23} \sin^2{2\theta_{13}} = \text{constant}$, 
is the outcome of the $\nu_{e}$ and $\bar{\nu}_{e}$ appearance 
measurement.
In Fig.~\ref{23fig1}b we combine the disappearance and 
appearance measurement and we end up with a pair of 
degenerate solutions of $\theta_{13}$ and $\theta_{23}$.

Now, we combine the information from reactor experiments. 
In a very good approximation what they measure is equal to  
the vacuum oscillation probability 
$P(\bar{\nu}_{e} \rightarrow \bar{\nu}_e)$ which is given 
again in the one $\Delta m^2$ dominance approximation as 
\begin{eqnarray}
P(\bar{\nu}_{e} \rightarrow \bar{\nu}_e) = 
1-\sin^2{2\theta_{13}} \sin^2 \left(\frac{\Delta m^2_{31} L}{4 E}\right). 
\label{Pvac_ee}
\end{eqnarray}
Given the value of $\Delta m^2_{31}$ by disappearance measurement, 
the reactor experiments determine $\theta_{13}$ independent 
of other mixing parameters.  This is illustrated in Fig.~\ref{23fig1}c. 
When we combine the accelerator disappearance and appearance 
experiments as well as reactor measurement, one of the two allowed 
solutions in Fig.~\ref{23fig1}b disappears as shown in Fig.~\ref{23fig1}d. 
Thus, the $\theta_{23}$ octant degeneracy can be lifted.

Though using the same method as proposed in \cite{MSYIS}, 
our analysis goes beyond that given in the reference in a number of ways. 
We have taken into account errors in accelerator disappearance 
and appearance as well as background so that the event number distributions 
roughly reproduce those obtained by the experimental group 
\cite{JPARC-detail,Hiraide}. 
Our treatment of the reactor measurement of $\theta_{13}$ is 
elaborated to include uncorrelated and correlated errors in order 
to treat the so called phase II type high statistics measurement.  
By using the highest sensitivity reactor and accelerator experiments, 
we aim at revealing the ultimate sensitivity for resolution 
of the $\theta_{23}$ octant degeneracy achievable by this method. 
Differences between this paper and  \cite{MSYIS} exist not only in 
the method for the analysis but also in some features of the results. 
For the particular set of parameters used in Fig.~\ref{23fig1}, 
the resolving power of the $\theta_{23}$ degeneracy is greater 
for the case with true values of $\theta_{23}$ in the first octant 
than for the case in the second octant. 
It is in disagreement with the naive expectation in \cite{MSYIS} 
based on the difference between $\theta_{23}$ of 
a true and a fake solutions. 
See Sec.~\ref{analysis_results} for more details.

\section{The $\theta_{23}$ octant degeneracy is hard to resolve 
by accelerator experiments}\label{robust}

We illustrate how difficult is to resolve the octant degeneracy by 
accelerator experiments with conventional neutrino beam. 
Apparently, this fact has been recognized by people in the community, 
but to our knowledge, 
a coherent discussion of why it is so has never been given. 
Therefore, we try to fill the gap. 
As a byproduct, the argument illuminates the nature of the 
$\theta_{23}$ degeneracy 
and may indicate a unique feature of our approach. 
We note that, strictly speaking, our following discussion in this section is valid
under circumstances that the matter effect can be treated
perturbatively, and hence for baseline shorter than $L \simeq 1000$ km.
(See below.)

\subsection{The octant degeneracy is robust}

Robustness of the degeneracy is obvious from 
Fig.~\ref{23fig1}a and Fig.~\ref{23fig1}b, 
but we want to elaborate the discussion 
to bring our understanding to a little deeper level. 
The reasons for the robustness are mainly twofold: 

\begin{itemize}

\item
The difference in energy spectra predicted by degenerate solutions 
in either the appearance or disappearance channels is too small to 
distinguish between the first- and the 
second-octant solutions.

\item
The matter effects in the disappearance channels are not strong 
enough to lift the octant degeneracy, unless one goes to very long baseline. 

\end{itemize}

\begin{figure}[htbp]
\vglue -1.5cm
\begin{center}
\hglue  1.0cm
\includegraphics[width=1.0\textwidth]{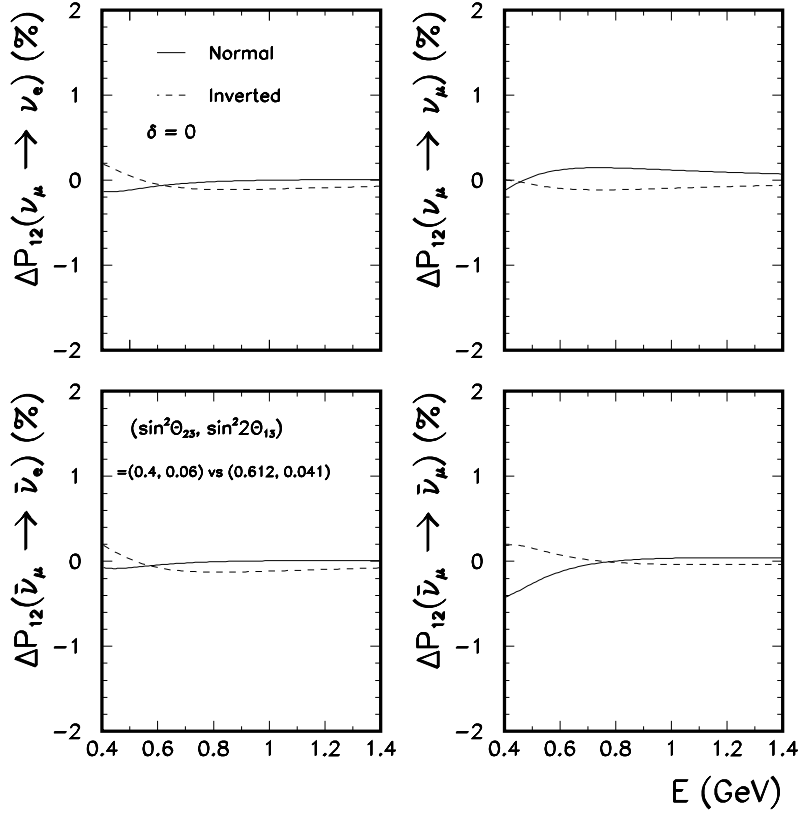}
\end{center}
\vglue -3.5cm
\caption{
Two examples of the $\theta_{23}$ degenerate solutions are shown to 
have almost the same energy spectrum in Kamioka with baseline of 
295 km from J-PARC. 
In the upper (lower) panels, we show the difference of 
the probabilities of two degenerate solutions for
neutrino (anti-neutrino) channels. 
The solid and the dashed lines are for normal ($\Delta m^2_{31} > 0$) 
and the inverted ($\Delta m^2_{31} < 0$) mass hierarchies, respectively. 
}
\label{app-disapp}
\end{figure}

To explain the first point, we introduce the quantity 
$\Delta P_{12}(\nu_{\alpha} \rightarrow \nu_{\beta})$, 
\begin{eqnarray}
\Delta P_{12}(\nu_{\alpha} \rightarrow \nu_{\beta}) \equiv
P(\nu_{\alpha} \rightarrow \nu_{\beta}; \theta_{23}^{\text{1st}},
\theta_{13}^{\text{1st}}) - P(\nu_{\alpha} \rightarrow \nu_{\beta};
\theta_{23}^{\text{2nd}}, \theta_{13}^{\text{2nd}}),
\label{deltaP}
\end{eqnarray}
the difference between probabilities with parameters of two degenerate 
solutions, $\theta_{23}^{i}$ and $\theta_{13}^{i}$ (i=1st, 2nd), 
determined at a particular value of energy in this exercise. 
In Fig.~\ref{app-disapp}, plotted are 
$\Delta P_{12}(\nu_{\mu} \rightarrow \nu_{e})$ (left upper panel),  
$\Delta P_{12}(\nu_{\mu} \rightarrow \nu_{\mu})$ (right upper panel), 
and their antineutrino counterparts in the lower two panels. 
As one can see in Fig.~\ref{app-disapp}, $\Delta P_{12}$'s are less
than $\sim 0.2$\% in most of the energy region.  In this exercise, the
CP phase $\delta^{\text{i}}$ (i=1st, 2nd) is kept equal and we did not
try to adjust it to make $\Delta P_{12}$ smaller.
We have also examined three other values of delta, 
$\delta=\pi/2, \pi, 3\pi/2$, in addition to the $\delta=0$ case in 
Fig.~\ref{app-disapp}, and reached the same conclusion. 

The smallness of 
$\Delta P_{12}(\nu_{\mu} \rightarrow \nu_{e})$ and   
$\Delta P_{12}(\nu_{\mu} \rightarrow \nu_{\mu})$ discourages the 
possibility of resolution of the $\theta_{23}$ octant degeneracy 
by using spectrum informations.  
This is in sharp contrast to the case of intrinsic degeneracy of 
$\theta_{13}$-$\delta$, for which the spectrum analysis is proved 
to be powerful in the presently discussed original T2K setting 
provided that $\theta_{13}$ is not too small \cite{T2KK}.


\subsection{Approximate analytic treatment of the $\theta_{23}$ octant degeneracy}
\label{analytic}

We present an analytic framework to understand better the reasons 
for the robustness of the $\theta_{23}$ octant degeneracy. 
We restrict ourselves to baselines where the earth matter effect 
can be treated as perturbation in the appearance channel. 
For a consistent treatment we start from the expression of 
the disappearance probability $P(\nu_{\mu} \rightarrow \nu_{\mu})$ 
in the one $\Delta m^2$ dominant vacuum oscillation approximation 
but with leading order $s^2_{13}$ correction\footnote{
Notice that the solar oscillation effect which is ignored in 
(\ref{Pvac_mumu2}) is small because of the suppression by a factor of 
$\sim (\Delta m^2_{21} / \Delta m^2_{31})^2$ at 
around the first oscillation maximum. 
}
\begin{eqnarray}
1 - P(\nu_{\mu} \rightarrow \nu_{\mu})  = 
\left[ 
\sin^2 2\theta_{23}
+ 4 s^2_{13} s^2_{23} 
\left(2 s^2_{23} - 1 \right)
\right]
\sin^2 \left(\frac{\Delta m^2_{31}L}{4E}\right). 
\label{Pvac_mumu2}
\end{eqnarray}
Then, a disappearance measurement determines $s^2_{23}$ 
to first order in $s^2_{13}$ as 
$(s^2_{23})^{(1)}= (s^2_{23})^{(0)} (1+ s^2_{13})$ 
where 
$(s^2_{23})^{(0)}$ is the solution obtained by ignoring $s^2_{13}$. 
By reading off the coefficient of sine squared term in (\ref{Pvac_mumu2}), 
one can determine $\sin^2{2\theta_{23}}$ and $s^2_{23}$ is given by 
$(s^2_{23})^{(0)}= \frac{1}{2} \left[ 1 \pm \sqrt{1- \sin^2{2\theta_{23}}} \right]$ 
as noted before. 
We note that the linear dependence of $(s^2_{23})^{(1)}$ 
on $s^2_{13}$ is clearly seen in Fig.~\ref{23fig1}a.

For the appearance channel, we use the $\nu_{e}$ ($\bar \nu_e$) 
appearance probability with first-order matter effect \cite{AKS}
\begin{eqnarray}
P[\nu_{\mu}(\bar{\nu}_{\mu}) &\rightarrow&
\nu_{\rm e}(\bar{\nu}_e)]  
\nonumber \\
&=&
\sin^2{2\theta_{13}} s^2_{23}
\left[
\sin^2 \left(\frac{\Delta m^2_{31} L}{4 E}\right)
-\frac {1}{2}
s^2_{12}
\left(\frac{\Delta m^2_{21} L}{2 E}\right)
\sin \left(\frac{\Delta m^2_{31} L}{2 E}\right) 
\right.
\nonumber \\
&&\hspace*{22mm} {}\pm
\left.
\left(\frac {4 Ea(x)}{\Delta m^2_{31}}\right)
\sin^2 {\left(\frac{\Delta m^2_{31} L}{4 E}\right)}
\mp 
\frac{a(x)L}{2}
\sin \left(\frac{\Delta m^2_{31} L}{2 E}\right) 
\right]
\nonumber \\
&&+
2J_{r} \left(\frac{\Delta m^2_{21} L}{2 E} \right)
\left[
\cos{\delta}
\sin \left(\frac{\Delta m^2_{31} L}{2 E}\right) \mp 
2 \sin{\delta}
\sin^2 \left(\frac{\Delta m^2_{31} L}{4 E}\right) 
\right]. 
\label{Pmue}
\end{eqnarray}
In (\ref{Pmue}), $a(x)= \sqrt 2 G_F N_e(x)$  \cite{MSW} 
where $G_F$ is the Fermi constant, $N_e(x)$ denotes the
electron number density at the point $x$ in the earth,
$J_r$ $(= c_{12} s_{12} c_{13}^2 s_{13} c_{23} s_{23} )$ 
denotes the reduced Jarlskog factor, and the upper and the 
lower sign $\pm$ refer to the neutrino and 
anti-neutrino channels, respectively. 

We make an approximation 
of ignoring terms of order 
$(\Delta m^2_{21}/\Delta m^2_{31}) J_r \cos{2\theta_{23}}$.  
Note that keeping only the leading order in this quantity 
is reasonable because $J_r < 0.04$, 
$\Delta m^2_{21}/\Delta m^2_{31} \simeq 1/30$, and 
$\cos{2\theta_{23}}= \pm 0.2$ for $\sin^2{2\theta_{23}} = 0.96$. 
Then, the two degenerate solutions obey an approximate relationship 
\begin{eqnarray}
\left( \sin^2 {2\theta_{13}} s^2_{23}  \right)^{1st} =
\left( \sin^2 {2\theta_{13}} s^2_{23}  \right)^{2nd}, 
\label{1st_order}
\end{eqnarray}
or, 
$s_{13}^{1st}s_{23}^{1st} = s_{13}^{2nd}s_{23}^{2nd}$
ignoring higher order terms in $s_{13}$. 
We can neglect the leading order correction in $s^2_{13}$ to 
$s^2_{23}$ in these relations because it gives $O(s^4_{13})$ terms.

\subsection{Appearance channel; $\nu_{\mu} \rightarrow \nu_{e}$}

With our machinery well oiled we can understand better the 
behavior of 
$\Delta P_{12}(\nu_{\mu} \rightarrow \nu_{e})$ and 
$\Delta P_{12}(\nu_{\mu} \rightarrow \nu_{\mu})$ given in 
Fig.~\ref{app-disapp}, in particular their small values, $\sim 10^{-3}$. 
We also note that even if we switch off the matter effect, there is 
no visible change in 
$\Delta P_{12}(\nu_{\mu} \rightarrow \nu_{e})$ and 
$\Delta P_{12}(\bar{\nu}_{\mu} \rightarrow \bar{\nu}_{e})$ 
in Fig.~\ref{app-disapp}. 
Now, these features can be understood in our analytic framework. 

Noting that 
$J_r^{1st} - J_r^{2nd} = \cos{2\theta_{23}^{1st}} J_r^{1st}$ 
in leading order in $\cos{2\theta_{23}}$, 
the difference between probabilities with the first and the second octant 
solutions can be given by 
\begin{eqnarray}
&&\Delta P_{12}(\nu_{\mu} \rightarrow \nu_{e}) 
\nonumber \\
&=&
2J_{r}^{1st} \cos{2\theta_{23}^{1st}} 
\left(\frac{\Delta m^2_{21} L}{2 E} \right)
\left[
\cos{\delta}
\sin \left(\frac{\Delta m^2_{31} L}{2 E}\right) \mp 
2 \sin{\delta}
\sin^2 \left(\frac{\Delta m^2_{31} L}{4 E}\right) 
\right]. 
\label{DeltaPmue}
\end{eqnarray}
The size of $\Delta P_{12}$ is about $\simeq 10^{-3}$ for 
$\cos{2\theta_{23}}= 0.2$, which is roughly consistent with the 
results given in Fig.~\ref{app-disapp}. 
The remarkable feature of (\ref{DeltaPmue}) is that the leading-order 
matter effect terms drops out completely, because they depend upon 
$\theta_{23}$ and $\theta_{13}$ only though the invariant 
combination (\ref{1st_order}). 
Therefore, we suspect that our statement about robustness of the 
$\theta_{23}$ octant degeneracy may apply not only to the 
T2K experiment but also to experiments with longer baseline, including
the NO$\nu$A project \cite{NOVA}.

\subsection{Disappearance channel; $\nu_{\mu} \rightarrow \nu_{\mu}$}

Does the disappearance channel $\nu_{\mu} \rightarrow \nu_{\mu}$ help?
Using the approximate relationship (\ref{1st_order}) and 
the first order (in $s^2_{13}$) corrected formula for $s^2_{23}$ 
mentioned after (\ref{Pvac_mumu2}), 
we obtain the expression of 
$\Delta P_{12}(\nu_{\mu} \rightarrow \nu_{\mu})$ 
to first order in matter effect as a sum of 
the vacuum and the matter effect contributions, 
%
$\Delta P_{12}(\nu_{\mu} \rightarrow \nu_{\mu}) = 
\Delta P_{12}(\nu_{\mu} \rightarrow \nu_{\mu})_{\text{vac}} + 
\Delta P_{12}(\nu_{\mu} \rightarrow \nu_{\mu})_{\text{matter}}$. 
%
The vacuum term
\begin{eqnarray}
\Delta P_{12}(\nu_{\mu} &\rightarrow& \nu_{\mu})_{\text{vac}} = 
\nonumber \\
&&
2 \cos 2 \theta_{23}^{\text{1st}}
\left[
-  \left(  s^2_{13} s^2_{23}   \right)^{\text{1st}}  
+  \sqrt{2}  s_{23}^{\text{1st}} J_r^{\text{1st}}  \cos\delta
\right]
\left(\frac{\Delta m^2_{21}L}{2E}\right) 
\sin \left(\frac{\Delta m^2_{31}L}{2E}\right)
\label{DeltaPmumu_vac} 
\end{eqnarray}
is small, $\sim 10^{-3}$, because of the suppression by either one of 
$s^2_{13}$ or $J_r$, and $\Delta m^2_{21} / \Delta m^2_{31}$.

The matter term is given by \cite{AKS} 
\begin{eqnarray}
\Delta P_{12}(\nu_{\mu} &\rightarrow& \nu_{\mu})_{\text{matter}} = 
2 P(\nu_{\mu} \rightarrow \nu_{\mu})_{\text{matter}} = 
2 (aL) \sin^22\theta_{13} s_{23}^2  D_{23} F(x), 
\label{DeltaPmumu_matter}
\end{eqnarray}
where 
$D_{23} \equiv \frac{1}{2} - s^2_{23}$, 
$x \equiv \frac{\Delta m^2_{31}L}{2E}$ 
and the function $F$ is defined as 
$F(x) \equiv \frac{4}{x} \sin^2\left(\frac{x}{2}\right) - \sin x$.  
Since the term flips sign under the  transformation 
$\theta_{23} \rightarrow \pi/2 - \theta_{23}$, it may be used to 
discriminate if $\theta_{23} < \pi/4$ or 
$\theta_{23} > \pi/4$ \cite{choubey}.
Unfortunately, the size of  (\ref{DeltaPmumu_matter}) 
and hence its contribution to 
$\Delta P_{12}(\nu_{\mu} \rightarrow \nu_{\mu})$ is small, 
\begin{eqnarray}
(\Delta P_{12})_{\text{matter}} \simeq
5.2  \times 10^{-3} F(x)
\left(\frac{\sin^22\theta_{13}}{0.1}\right)
\left(\frac{D_{23}}{0.1}\right)
\left(\frac{\rho}{2.8 \mbox{ gcm}^{-3}}\right)
\left(\frac{L}{1000 \mbox{ km}}\right). 
\label{magPmatt}
\end{eqnarray}
Note that $F$ is a monotonically increasing 
function of $x$ in $0 < x \lsim 4$, and $F(\pi) = 4/\pi = 1.273$.
Thus, the difference $(\Delta P_{12})_{\text{matter}}$ is 
of order $10^{-3}$ at $L=300$ km 
(less than 1\% even at $L=1000$ km) with $\theta_{13}$ 
at around the Chooz limit \cite{CHOOZ}. 

It should be noticed that 
$\Delta P_{12}$ flips sign not only under 
$D_{23} \rightarrow -D_{23}$ 
but also under 
$\Delta m^2_{31} \rightarrow -\Delta m^2_{31}$. 
Therefore, measurement of the sign of $\Delta P_{12}$ determines 
the combined sign, $\Delta m^2_{31} \times D_{23}$.
Hence, the use of this effect to resolve the $\theta_{23}$ degeneracy 
requires prior knowledge of the neutrino mass hierarchy,  i.e.,
the sign of $\Delta m^2_{31}$.

Thus, $\Delta P_{12}$ is small, typically 0.1\% level in the T2K setting, 
both in the appearance and the disappearance channels. 
This feature explains well the behavior shown in Fig.~\ref{app-disapp}.

\subsection{Approximate decoupling of the $\theta_{23}$ octant degeneracy} 
\label{decoupling}

Under the approximation of ignoring $\Delta P_{12}$ 
in the appearance and the disappearance channels, 
there is no way to resolve the $\theta_{23}$ degeneracy by 
spectrum analysis. 
But, on the other hand, it implies that the octant degeneracy 
decouples from the other types of degeneracies, 
$\Delta m^2_{31}$-sign and the intrinsic ones. 
The desirable feature prevails after the $\theta_{23}$ degeneracy is 
resolved because it is executed by combining the reactor 
measurement of $\theta_{13}$ in our method, 
which is free from any degeneracies. 
We cannot lift the $\Delta m^2_{31}$-sign degeneracy by the method 
we explore in this paper, but we can resolve the $\theta_{23}$ octant 
degeneracy independent of the sign of $\Delta m^2_{31}$. 
We will explicitly verify this point in Sec.~\ref{analysis_results} 
by performing the analysis under assumptions of the right and 
the wrong mass hierarchies.

A simple remark on the intrinsic degeneracy of $\theta_{13}$-$\delta$; 
The coupling between the $\theta_{23}$ and the intrinsic degeneracy 
can be avoided by doing measurement at the oscillation maximum, 
or more precisely the ``thinnest ellipse'' limit \cite{KMN02}.
Or, if necessary, it can be resolved relatively easily 
(compared to the $\Delta m^2_{31}$-sign degeneracy) by doing 
the spectrum analysis, as demonstrated in the T2K setting in 
\cite{T2KK}.

\section{analysis method}
\label{analysis_method}

In this section, we summarize the statistical method and the 
procedure of our analysis. 
We ask the readers to refer to Appenix \ref{number} for any details of 
how the numbers of events are computed. 
We use the full three-flavor oscillation formulas in our analysis. 
They are obtained by numerically solving the neutrino 
evolution equation with the constant electron number density.

For concreteness, we consider for accelerator experiment 
the phase II of the T2K project~\cite{JPARC}
where the beam power is upgraded to 4 MW and the far detector will be 
Hyper-Kamiokande with 0.54 Mt fiducial volume.  
We consider the 2.5 degree off axis $\nu_\mu$ beam which 
has a peak at around 0.65 GeV. 
We assume the exposures of 2 and 6 years of neutrino and 
anti-neutrino running, respectively \cite{JPARC}.   
For the reactor experiment, we consider the exposure 
of 10 GW$\cdot$kt$\cdot$yr.

\subsection{$\nu_\mu \to \nu_e$ ($\bar{\nu}_\mu \to \bar{\nu}_e$) appearance mode}

The $\nu_\mu \to \nu_e$ ($\bar{\nu}_\mu \to \bar{\nu}_e$) 
appearance mode is important to determine precisely 
the value of $\sin^2\theta_{23} \sin^2 2\theta_{13}$, 
provided that $\Delta m_{23}^2$ is well determined, 
which is possible by the disappearance mode. 
As discussed in Ref.~\cite{KMN02}, if the experiment is done 
at or close to the oscillation maximum, we can determine well 
the quantity $\sin^2\theta_{23} \sin^2 2\theta_{13}$, even if
we do not know the value of the CP phase $\delta$. 
Since information on the energy dependence of this channel 
cannot be important in resolving $\theta_{23}$ degeneracy 
(as we saw in the previous section), 
we consider only the
total number of events and define $\chi^2$ for 
the appearance channel as follows,
\begin{equation}
\chi_{\text{app}}^2 \equiv 
\frac{ (N_{\rm sig}^{\text{obs}}+N_{\text{\tiny BG}}^{\text{obs}} 
-N_{\rm sig}^{\text{theo}}-N_{\text{\tiny BG}}^{\text{theo}} 
)^2}
{N_{\rm sig}^{\text{obs}}+N_{\text{\tiny BG}}^{\text{obs}} 
+ (\sigma_{\text{sig}}N_{\rm sig}^{\text{obs}})^2
+ (\sigma_{\text{\tiny BG}}N_{\text{\tiny BG}}^{\text{obs}})^2}, 
\label{chi2_app}
\end{equation}
where $N^{\text{obs}}$ and $N^{\text{theo}}$ are 
the number of events to be observed and the theoretically expected  
one, respectively, for given values of the oscillation parameters. 
We note that background events come mainly from 
neutral current interactions as well as the events induced by
$\nu_e$ ($\bar{\nu}_e$) which is inevitably contained in the initial flux. 
We assume optimistic systematic errors, 
$\sigma_{\text{sig}}=\sigma_{\text{\tiny BG}}=2$ \%
for the T2K II experiment.

\begin{table}[th]
\caption[aaa]{Number of events in the appearance mode with and without
oscillation. We assume 2 (6) years of neutrino (anti-neutrino) running
with the T2K phase II set up.
For the case with oscillation,  we set
$\Delta m^2_{31} = 2.5\times 10^{-3}$ eV$^2$,
$\Delta m^2_{21} = 0$, 
$\sin^2 2\theta_{23}=1$, and $\sin^2 2\theta_{13}=0.1$, with
the matter density $\rho = 2.3$ g/cm$^3$ and the electron
fraction (number of electron per nucleon) being $Y_e$=0.5.
Numbers in parentheses correspond to the case where the
matter effect is switched off.
}
\vglue 0.5cm
\begin{tabular}{|c||c||c||c|}
\hline
 Case 1 ($\nu_{\mu} \rightarrow \nu_e$) & Signal events   & BG NC events
&  BG beam events  \\
\hline
No Oscillation  &   0     &  542   &  785          \\
\hline
Oscillation  & 8683 (8016)     & 542   & 724 (728)   \\
\hline \hline
 Case 2 ($\bar{\nu}_{\mu} \rightarrow \bar{\nu}_{e}$) &
&   &    \\
\hline \hline
No Oscillation    & 0       & 624    &  817          \\
\hline
Oscillation       &   7340 (7990)   &  624   & 761 (757)          \\
\hline
\end{tabular}
\label{Tab-events}
\vglue 0.5cm
\end{table}

In Table I, we show the expected number of events for the T2K phase II 
for the 2 (6) years of exposure for neutrino (anti-neutrino) running
with and without oscillation effect. 
For the case with oscillation, we assumed the oscillation parameters 
$\Delta m^2_{31} = 2.5\times 10^{-3}$ eV$^2$, 
$\Delta m^2_{21} = 0$, 
$\sin^2 2\theta_{23}=1$, and 
$\sin^2 2\theta_{13}=0.1$. 
The choice is to compare the number of events to the one quoted 
in Ref.~\cite{JPARC-detail} for T2K I after properly scaling the 
fiducial volume, exposure time and the beam power. 
We have confirmed that our results agree reasonably well 
with the numbers quoted in \cite{JPARC-detail}. 
%

\begin{figure}[htbp]
\vglue -2.5cm
\begin{center}
\includegraphics[width=0.8\textwidth]{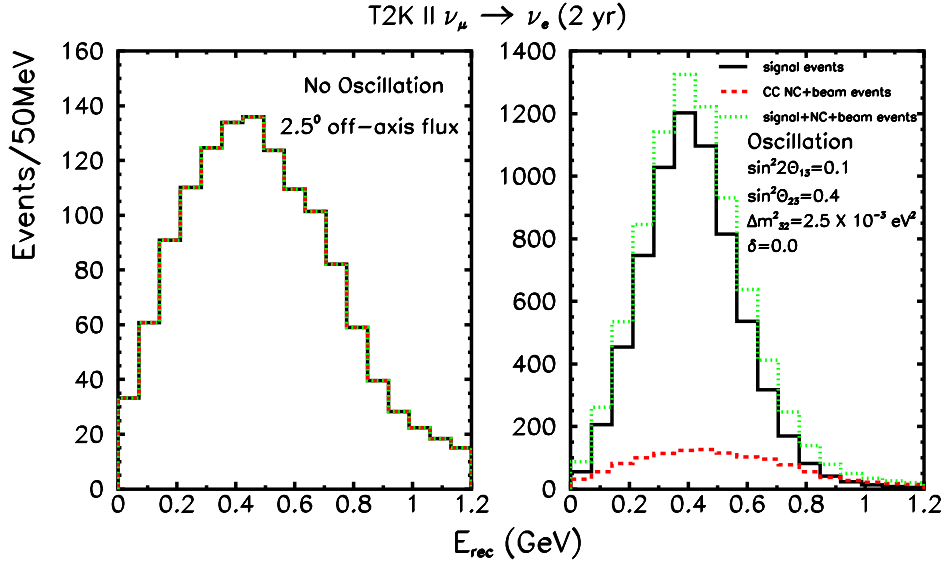}
\end{center}
\vglue -2.5cm
\caption{
Examples of event number distributions are plotted as a function of the 
reconstructed neutrino energy for the appearance mode
with and without oscillations. 
In the latter, the following values of the oscillation parameters are used: 
$\Delta m^2_{31} = 2.5\times 10^{-3}$ eV$^2$, 
$\sin^2 \theta_{23}=0.4$, $\sin^2 2 \theta_{13}=0.1$, 
$\Delta m^2_{21} = 8.0\times 10^{-5}$ eV$^2$, 
$\sin^2 \theta_{12}$ = 0.31, $\delta = 0$. 
}
\label{app_spectrum}
\end{figure}

Although we do not use the spectrum informations in our analysis, we
present in Fig.~\ref{app_spectrum} some examples of the energy
distribution of the $\nu_{e}$ appearance events for completeness.  For
this plot, we set the oscillation parameter $\Delta m^2_{31} =
2.5\times 10^{-3}$ eV$^2$, $\sin^2 \theta_{23}=0.4$, $\sin^2 2
\theta_{13}=0.1$, $\Delta m^2_{21} = 8.0\times 10^{-5}$ eV$^2$,
$\sin^2 \theta_{12}$ = 0.31, $\delta = 0$.
Throughout our analysis, we fix the solar neutrino mixing 
parameters $\Delta m^2_{21}$ and $\sin^2 \theta_{12}$ to 
these values. 
One can observe in the figure that background is quite small in size 
and has similar shape as the signal events. 
Notice also that the modulation of energy spectrum by neutrino 
oscillation is rather modest. 

\subsection{$\nu_\mu \to \nu_\mu$ 
($\bar{\nu}_\mu \to \bar{\nu}_\mu$) disappearance mode}

\begin{figure}[htbp]
\vglue -1.5cm
\begin{center}
\includegraphics[width=0.8\textwidth]{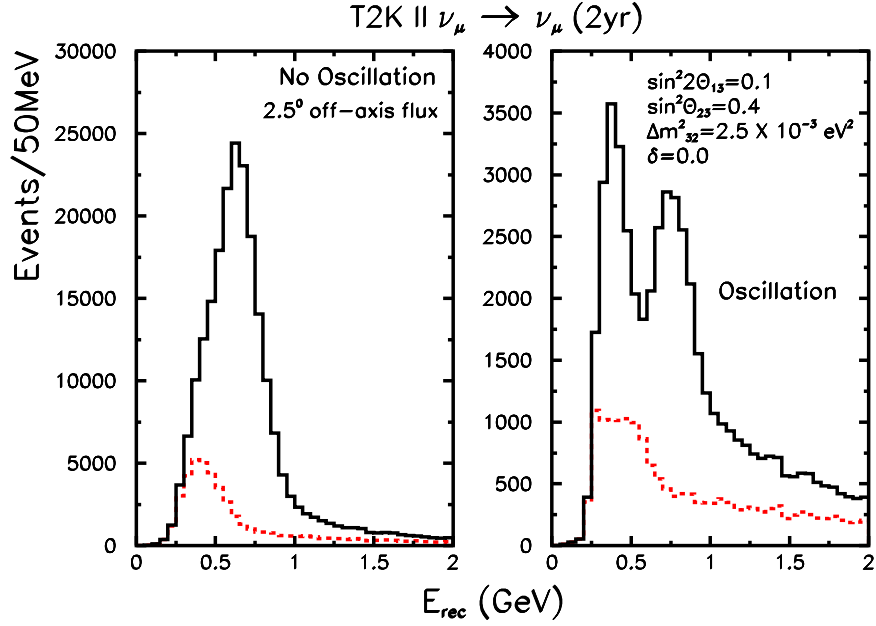}
\end{center}
\vglue -3.8cm
\caption{
Examples of event number distributions are plotted as a function of the 
reconstructed neutrino energy for the disappearance mode
without oscillation (left panel) and with oscillation (right 
panel) with the mixing parameters 
$\Delta m^2_{31} = 2.5\times 10^{-3}$ eV$^2$, 
$\sin^2 \theta_{23}=0.4$, $\sin^2 2 \theta_{13}=0.1$, 
$\Delta m^2_{21} = 8.0\times 10^{-5}$ eV$^2$, 
$\sin^2 \theta_{12}$ = 0.31 and $\delta = 0$. 
The histogram by the solid lines indicate 
the sum of the signal and background events whereas the ones 
by the dashed line indicate only the background events.  
}
\label{event_distribution}
\end{figure}

The $\nu_\mu \to \nu_\mu$ ($\bar{\nu}_\mu \to \bar{\nu}_\mu$) 
disappearance mode is important to determine 
$\sin^2 2\theta_{23}$ as well as $\Delta m^2_{31}$ accurately. 
In Fig.~\ref{event_distribution}, we show the expected event 
number distribution as a function of the reconstructed neutrino energy 
in the absence (left panel) and 
in the presence (right panel) of oscillation 
with the mixing parameters, $\sin^2 \theta_{23}=0.4$, 
 $\Delta m^2_{23}=2.5\times 10^{-3}$ eV$^2$, 
$\sin^2 2\theta_{13}=0.1$, 
and $\delta = 0$.  
Unlike the case of the appearance channel, the energy distribution 
is significantly modified by the oscillation effect.

For the disappearance mode, we consider 36 bins with 50 MeV width 
from 0.2 GeV to 2.0 GeV in terms of the reconstructed neutrino energy.  
The $\chi^2$ function is defined as follows, 

\begin{equation}
\chi_{\text{dis}}^2 
\equiv \min_{\alpha_{\text{sig}}, \alpha_{\text{BG}}}
\sum_i
\frac{ [N_i^{\text{obs}} + N_{i,\text{BG}}^{\text{obs}} 
-(1+\alpha_{\text{sig}})   N_i^{\text{theo}}
-(1+\alpha_{\text{BG}})    N_{i,\text{BG}}^{\text{theo}}
]^2}
{N_i^{\text{obs}}+N_{i,\text{BG}}^{\text{obs}} }
+ \left(\frac{\alpha_{\text{sig}}}{\sigma_{\text{sig}}}\right)^2
+ \left(\frac{\alpha_{\text{BG}}}{\sigma_{\text{BG}}}\right)^2
, 
\label{chi2_dis}
\end{equation}
where $N_{\text{sig},i}^{\text{obs}}$ and $N_{\text{sig},i}^{\text{theo}}$ 
are, the number of signal events to be observed and 
the theoretically expected  
one, respectively for the $i$-th bin, 
and $N_{\text{\tiny BG},i}^{\text{obs}}$ and 
$N_{\text{\tiny BG},i}^{\text{theo}}$ are the 
corresponding background event numbers.
We assume the optimistic values for the systematic errors, 
$\sigma_{\text{sig}}=\sigma_{\text{BG}}=2$\%
for the T2K phase II. 

\begin{figure}[htbp]
\vglue 1.0cm
\begin{center}
\includegraphics[width=0.8\textwidth]{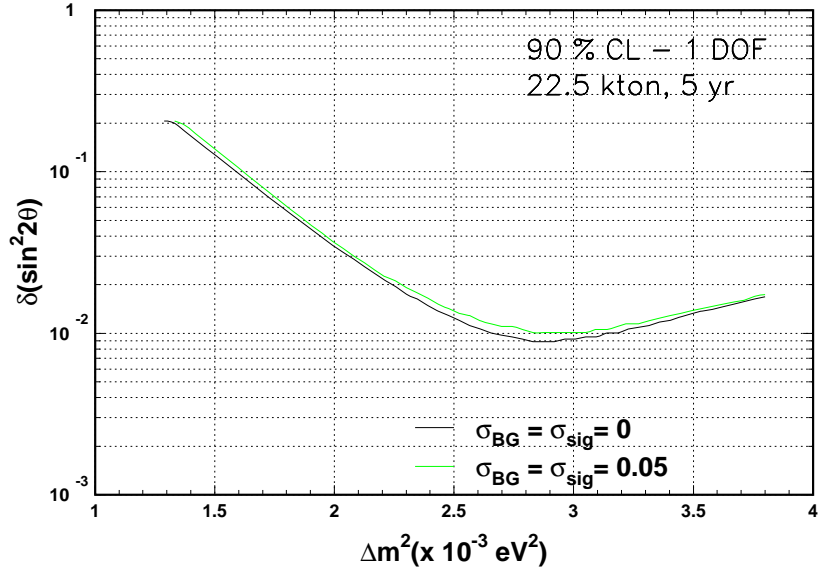}
\end{center}
\vglue -7.0cm
\caption{
Expected sensitivity on $\sin^22\theta_{23}$ as a function of 
the true value of $\Delta m^2_{23}$ obtained with the $\chi^2$ 
defined in (\ref{chi2_dis}) assuming the pure 2 flavor oscillation. 
}
\label{Fig:sensitivity_angle}
\end{figure}

In Fig.\ref{Fig:sensitivity_angle}, we show the expected sensitivity 
for $\sin^2 2\theta_{23}$ assuming the pure 2 flavor oscillation
($\theta_{13}=0$) as a function of 
the true value of $\Delta m_{23}^2$, which compares reasonably 
well with that of \cite{Hiraide}.
We have checked that this simple $\chi^2$ can reproduce the 
expected T2K sensitivity based on more realistic 
calculations obtained by Monte Carlo simulation given in Ref.~\cite{Hiraide}. 

\subsection{$\bar{\nu}_e \to \bar{\nu}_e$ disappearance mode}

For the reactor experiment, we use the same $\chi^2$ function used 
in our previous works \cite{reactorCP,sado}, which is defined as,
\begin{eqnarray}
 \chi^2_{\text{reac}}
 &\equiv&
 \min_{\text{$\alpha$'s}}
  \sum_{a = f, n}
          \left[
           \sum_{i=1}^{17}
            \left\{
             \frac{
              \left( N^{\rm theo}_{ai}
                 - ( 1 + \alpha_i + \alpha_a + \alpha ) N^{\rm obs}_{ai}
              \right)^2 }
              {   N^{\rm obs}_{ai}
                + \sigma_{\rm db}^2 (N^{\rm obs}_{ai})^2 }
             + \frac{ \alpha_i^2 }{ \sigma_{\rm Db}^2 }
            \right\}
           + \frac{ \alpha_a^2 }{ \sigma_{\rm dB}^2 }
          \right]
          + \frac{ \alpha^2 }{ \sigma_{\rm DB}^2 },
\label{chi2_reac}
\end{eqnarray}
where $N_{ai}^{\rm theo}$ represents the theoretical number of events
at a near ($a=n$) or a far ($a=f$) detector within the $i$-th bin whose
width is 0.425 MeV. 
We set the distance to the near and the far detectors to be 300 m and 1500 m,  
respectively. 
$N_{ai}^{\rm obs}$ are number of events to be observed. 
We consider four types of systematic error: $\sigma_{\rm DB}$,
$\sigma_{\rm Db}$, $\sigma_{\rm dB}$ and $\sigma_{\rm db}$. The 
subscript D (d) represents the fact that the error is correlated 
(uncorrelated) between detectors. 
 The subscript B (b) represents the fact that the error is correlated 
(uncorrelated) among bins. 
The $\alpha$'s are parameters to be varied freely in order to 
take into account these  systematic errors.
In this work, we consider the high precision reactor experiments 
whose sensitivity can go beyond the ones 
currently expected by experiments such as 
Double-Chooz~\cite{DC} and KASKA~\cite{Kaska}. 
Namely, we assume sensitivities below $\sin^2 2\theta_{13} = 0.01$, 
the one expected to be achievable by phase-II type 
experiments such as the 
Braidwood~\cite{braidwood}, 
the Daya Bay \cite{dayabay}, and 
the Angra~\cite{angra} projects.

Computation of the event number is done in the same way as in
Ref.~\cite{sado}.  We ignore the possible contribution from
geo-neutrinos.  It is demonstrated recently by KamLAND that its flux
is consistent with the one expected by geo-chemical earth models
\cite{KamLAND_geo}.  In this case, the effect of geo-neutrinos is
negligibly small at the baseline of $\sim$1 km, as one can easily
guess by extrapolation of the situation in a reactor $\theta_{12}$
experiment with baseline of $\sim$60 km \cite{sado}.

To characterize reactor measurement,
we use GW$\cdot$kt$\cdot$yr, the  ``total exposure'' unit, 
which is defined as the product of the net values of the reactor 
thermal power (in GW), detector fiducial volume (in kton) 
and running time (in year). 
We consider the reactor measurement for 10 GW$\cdot$kt$\cdot$yr. 
The total number of events at the far detector is $1.63 \times 10^{6}$. 
We consider two different sets of systematic errors:  
a relatively conservative choice and an optimistic one.
For the conservative choice, we adopt similar values  of the
systematic errors  used in Ref.~\cite{reactorCP,sado},
$\sigma_{\rm DB}=\sigma_{\rm Db}=2.0$ \% and 
$\sigma_{\rm dB}=\sigma_{\rm db}=0.5$ \%, 
and for the optimistic one, we set 
$\sigma_{\rm DB}=\sigma_{\rm Db}=1.0$ \% and 
$\sigma_{\rm dB}=$ 0.2\% and $\sigma_{\rm db}=0.2$ \%. 
The latter extremely small errors may be difficult to reach, but 
they are used to estimate the upper limit of resolving power of the 
$\theta_{23}$ degeneracy by the present method. 
The sensitivity limit of $\theta_{13}$ (assuming no depletion) at 
$\Delta m^2_{31}=2.5 \times 10^{-3}$ eV$^2$ is
$\sin^2 2\theta_{13} = 1.14 \times 10^{-2}$  ($2.63 \times 10^{-2}$) 
at 1$\sigma$ (3$\sigma$) CL
for the relatively conservative errors, and 
$\sin^2 2\theta_{13} = 5.39 \times 10^{-3}$  ($1.25 \times 10^{-2}$) 
at 1$\sigma$ (3$\sigma$) CL for the optimistic ones. 

\subsection{Combined analysis}

For the combined analysis, we simply sum all the $\chi^2$ functions 
defined in Eqs. (\ref{chi2_app}), (\ref{chi2_dis}) and (\ref{chi2_reac}), 
\begin{equation}
\chi^2 = \chi_{\text{app}}^2 + \chi_{\text{dis}}^2 + \chi_{\text{reac}}^2. 
\end{equation}
The allowed region in the 
$\sin^2 2\theta_{13}-\sin^2\theta_{23}$ plane 
is determined by the usual condition, 
$\Delta \chi^2 \equiv \chi^2-\chi^2_{\text{min}} < $ 
2.3, 6.18 and 11.83 for 1, 2 and 3 $\sigma$ CL for two degrees of freedom. 
We will establish the parameter regions where we can resolve the 
$\theta_{23}$ octant degeneracy for 1 degree of freedom by imposing
the condition $| \chi^2_\text{min}(\theta^{\text{true}}_{23}) -
\chi^2_\text{min}(\theta^{\text{false}}_{23})| > $ 2.71, 4 and 6.63
for 90, 95 and 99\% CL, respectively, where
$\theta^{\text{true}}_{23}$ and $\theta^{\text{false}}_{23}$ are,
respectively, the true and the false value of $\theta_{23}$.

\section{Analysis Results}
\label{analysis_results}

In this section, we show our results based on our analysis with the
combined $\chi^2$ of all channels. To remind the readers, our analysis
is based on the T2K II experiment of 2 (6) years running of neutrino
(anti-neutrino) modes with 4MW beam power with the Hyper-Kamiokande
detector whose fiducial volume is 0.54 Mt \cite{JPARC}.  For the
reactor experiment, the exposure of 10 GW$\cdot$kt$\cdot$yr is
assumed.

We assume throughout this section that the mass hierarchy 
(determined by nature) is normal type 
($\Delta m^2_{31}>0$) unless otherwise stated.  
Even if we repeat the same procedure with the true mass hierarchy 
of inverted type ($\Delta m^2_{31} < 0$), the region in which 
the degeneracy is solved is remarkably similar. 
Therefore, we decided to concentrate on the normal hierarchy case. 
Of course, we examine the stability of our results by assuming 
the wrong hierarchy in the analysis.

\begin{figure}[htbp]
\vglue -1.0cm
\begin{center}
\includegraphics[width=1.0\textwidth]{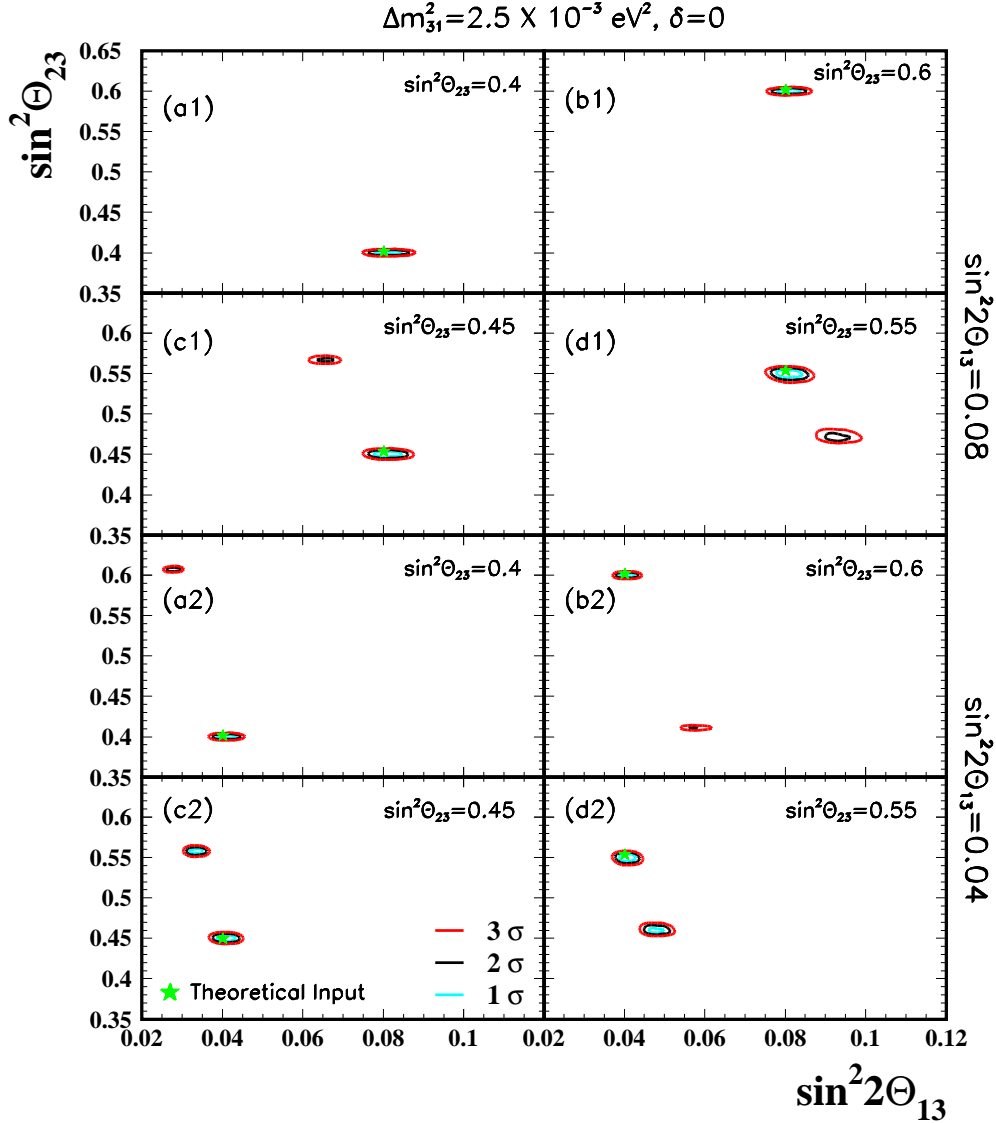}
\end{center}
\vglue -1.5cm
\caption{
Some examples of the allowed regions in the 
$\sin^2\theta_{13}-\sin^2\theta_{23}$ plane 
determined by the combined analysis 
for various different input parameters. 
For all plots, the input values of 
$\Delta m^2_{31} = 2.5 \times 10^{-3}$eV$^2$ and the CP phase $\delta$ = 0. 
The input value of $\theta_{13}$ for the upper (lower) four panels
is $\sin^2\theta_{23}=0.08$ (0.04). 
The input values of $\theta_{23}$ are 
(a) $\sin^2\theta_{23}=0.4$,  (b) $\sin^2\theta_{23}=0.6$,  
(c) $\sin^2\theta_{23}=0.45$ and (d) $\sin^2\theta_{23}=0.55$. 
The conservative values of the systematic errors are taken for 
reactor measurement; 
$\sigma_{\rm DB}=\sigma_{\rm Db}=2.0$ \% and 
$\sigma_{\rm dB}=\sigma_{\rm db}=0.5$ \%. 
}
\label{allowed_region1}
\end{figure}

\begin{figure}[htbp]
\vglue -1.0cm
\begin{center}
\includegraphics[width=1.0\textwidth]{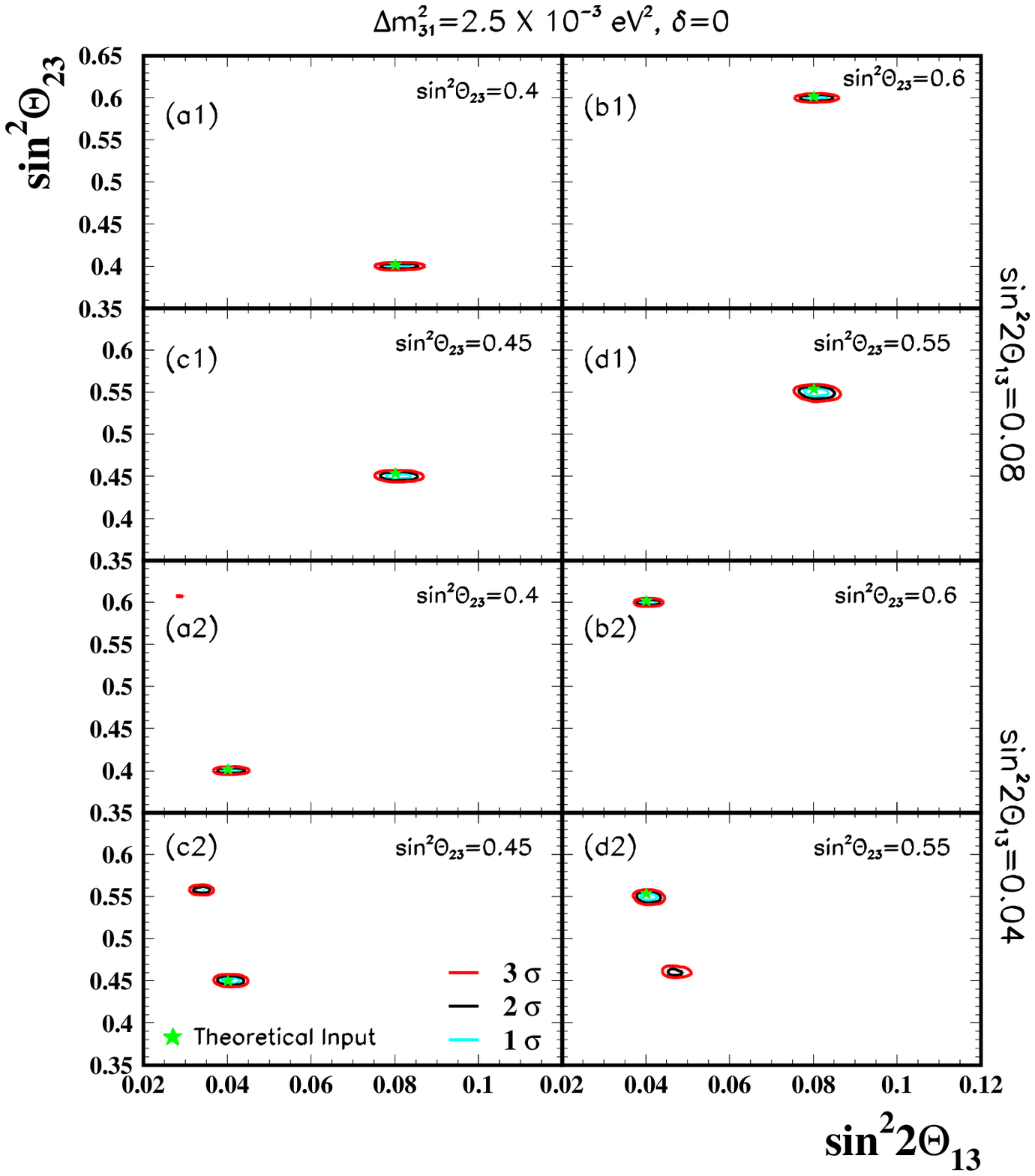}
\end{center}
\vglue -1.5cm
\caption{
Same as Fig.~\ref{allowed_region1} but for smaller systematic errors for 
the reactor experiment, 
$\sigma_{\text{DB}}=\sigma_{\text{Db}} = 1$ \%, 
$\sigma_{\text{db}}=0.2$ \% and $\sigma_{\text{dB}} = 0.2$ \%. 
}
\label{allowed_region2}
\end{figure}

In Fig.~\ref{allowed_region1}, we show the allowed regions in 
the $\sin^2 2\theta_{13} - \sin^2 \theta_{23}$ plane at 
1 $\sigma$ (light blue curve), 2 $\sigma$ (black curve), and 
3 $\sigma$ (red curve) CL 
for 2 degree of freedom,  
for 8 different sets of input parameters; 
$(\sin^2 2\theta_{13}, \sin^2 \theta_{23})$ = (0.08, 0.4), (0.08,
0.6), (0.08, 0.45), (0.08, 0.55), (0.04, 0.4), (0.04, 0.6) and (0.04,
0.45), (0.04, 0.55), which are indicated by the symbol of a star.
They are determined by the combined $\chi^2$ analysis using all the
channels, $\nu_\mu \to \nu_e$ ($\bar{\nu}_\mu \to \bar{\nu}_e$)
appearance mode, by $\nu_\mu \to \nu_\mu$ ($\bar{\nu}_\mu \to
\bar{\nu}_{\mu}$) disappearance mode and by $\bar{\nu}_e\to
\bar{\nu}_e$ disappearance mode.  The relatively conservative values
of the systematic errors are taken for the reactor measurement;
$\sigma_{\rm DB}=\sigma_{\rm Db}=2.0$ \% and $\sigma_{\rm
  dB}=\sigma_{\rm db}=0.5$ \%.
In our $\chi^2$ analysis, for given values of input parameters, 
we vary not only $\theta_{23}$ and $\theta_{13}$ but also 
$\delta$ and $\Delta m^2_{31}$, as these parameters should  
be determined by the fit. 
Note, however, that the range of 
$\Delta m^2_{31}$ and $\theta_{23}$ are restricted to the ones
constrained by atmospheric neutrino experiments~\cite{SKatm}.

For a larger value of $\sin^2 2\theta_{13} = 0.08$, 
the octant degeneracy is resolved for $\sin^2 2\theta_{23} = 0.96$ 
(see Fig.~\ref{allowed_region1}a1 and b1), 
whereas for $\sin^2 2\theta_{23} = 0.99$, 
the degeneracy can be resolved only at 1 $\sigma$ CL but not
at 2 $\sigma$ CL or higher (see Fig.~\ref{allowed_region1}c1 and d1).
For a smaller value of $\sin^2 2\theta_{13} = 0.04$, 
for $\sin^2 2\theta_{23} = 0.96$, the octant degeneracy 
is resolved only at 1 $\sigma$ CL 
(see Fig.~\ref{allowed_region1}a2 and b2)
and for $\sin^2 2\theta_{23} = 0.99$, 
the degeneracy can not be resolved 
even at 1 $\sigma$ CL (see Fig.~\ref{allowed_region1}c2 and d2).

In Fig.~\ref{allowed_region2}, we show the same quantities 
but for the small systematic errors for reactor experiments, 
$\sigma_{\text{DB}}=\sigma_{\text{Db}} = 1$ \%, 
$\sigma_{\text{db}}=\sigma_{\text{dB}} = 0.2$ \%, 
the very optimistic ones.  
In this case, the octant degeneracy is completely resolved 
for a large $\theta_{13}$, $\sin^2 2\theta_{13} = 0.08$ both for 
$\sin^2 2\theta_{23} = 0.96$ and 0.99 
(see Fig.~\ref{allowed_region2}a1, b1, c1, and d1).
For a small value of $\theta_{13}$, $\sin^2 2\theta_{13} = 0.04$, 
however, we have a mixed result; 
For $\sin^2 2\theta_{23} = 0.96$, the $\theta_{23}$ 
degeneracy is resolved both for $\sin^2 2\theta_{23} = 0.96$ and 0.99 
apart from a tiny 3$\sigma$ CL region in 
Fig.~\ref{allowed_region2}a2. 
For $\sin^2 2\theta_{23} = 0.99$, the degeneracy is not lifted, except in 
Fig.~\ref{allowed_region2}d2 where the clone solution disappears at 
1$\sigma$ CL.

\begin{figure}[htbp]
\vglue -2.5cm
\begin{center}
\includegraphics[width=1.0\textwidth]{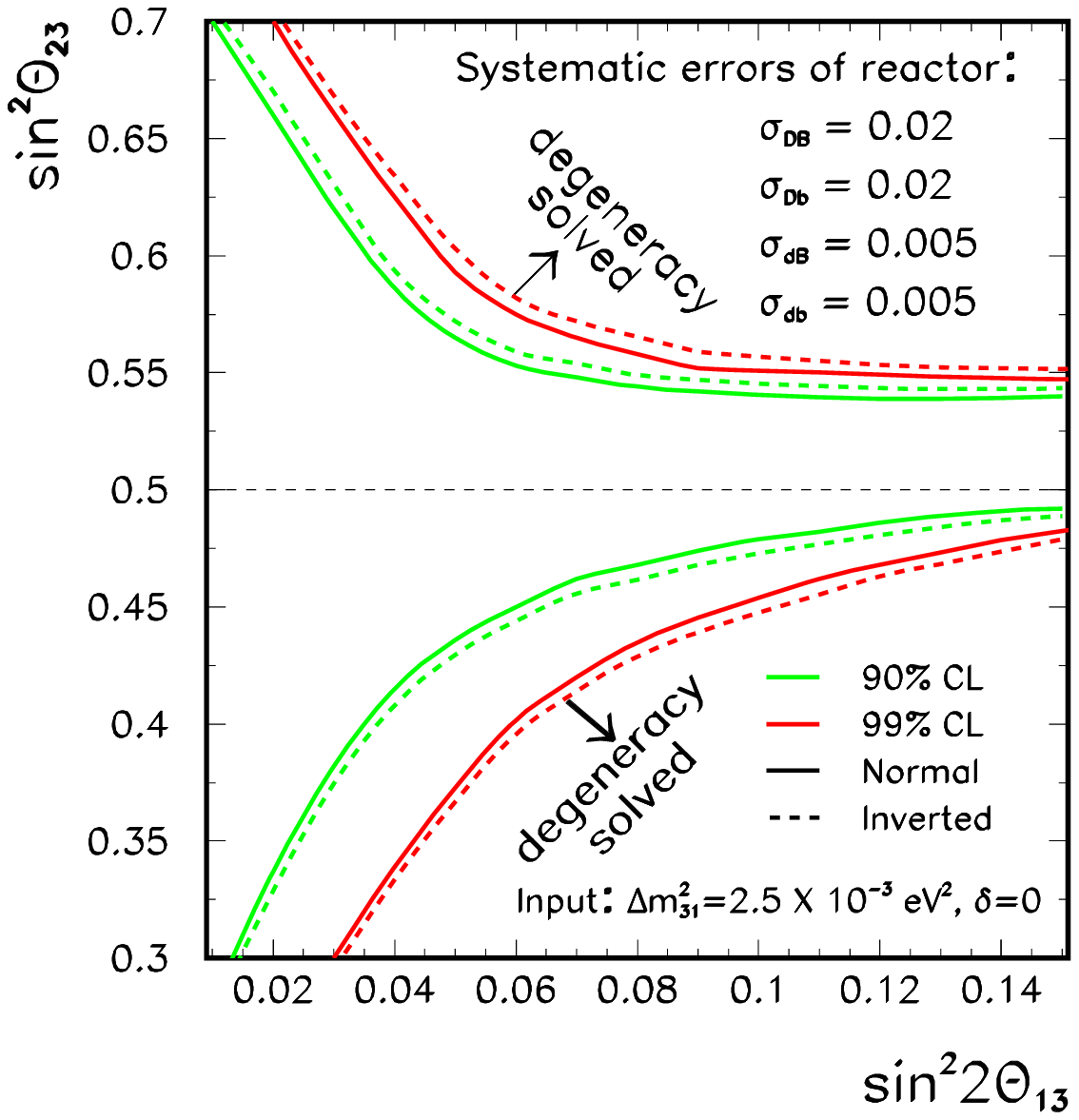}
\end{center}
\vglue -4.5cm
\caption{
The region in $\sin^2 2\theta_{13} - \sin^2 \theta_{23}$ space 
where the $\theta_{23}$ octant degeneracy can be resolved 
at 90\% (thin green) and 99\% (thick red) CL. 
The solid (dashed) curve is for the case of taking the normal (inverted) 
hierarchy to perform the fit, assuming the normal hierarchy as input. 
Conservative systematic errors, as indicated in the figure, are
considered here.
}
\label{region_resolved1}
\end{figure}

\begin{figure}[htbp]
\vglue -2.5cm
\begin{center}
\includegraphics[width=1.0\textwidth]{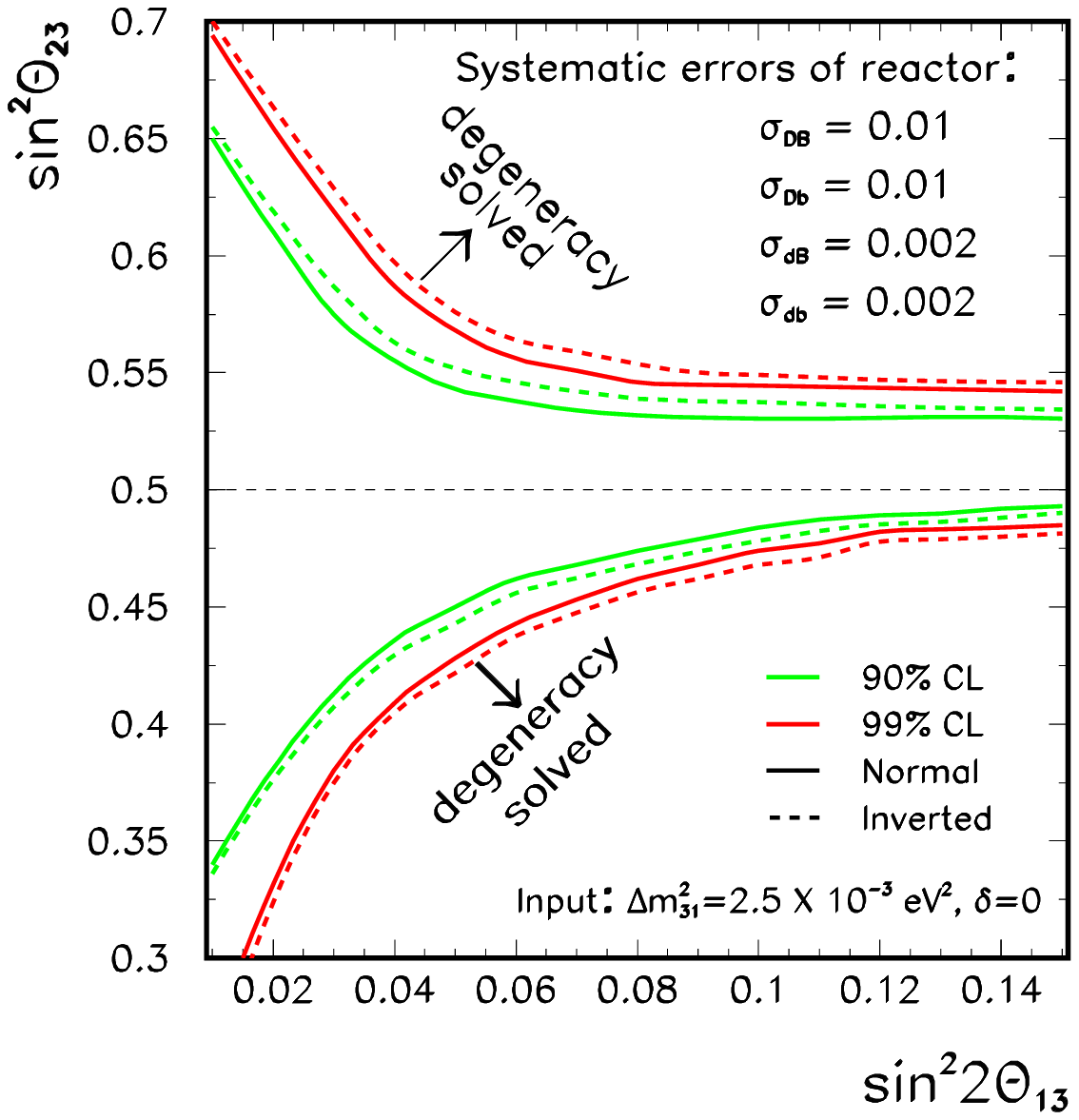}
\end{center}
\vglue -4.5cm
\caption{
The same as in Fig.~\ref{region_resolved1} but with 
optimistic systematic errors. 
}
\label{region_resolved2}
\end{figure}


In Fig.~\ref{region_resolved1} and Fig.~\ref{region_resolved2}, 
we show the region bounded by the solid and the dashed curves 
in which we can resolve the $\theta_{23}$ octant degeneracy in the 
$\sin^2\theta_{13}$-$\sin^2\theta_{23}$ plane at 
90\% (thin green curves) and 99\% (thick red curves) CL 
for 1 degree of freedom for the conservative and the optimistic 
values of the reactor systematic errors, respectively. 
Namely, the region inside the bands correspond to the parameters 
which satisfies 
$| \chi^2_\text{min}(\theta^{\text{true}}_{23})
- \chi^2_\text{min}(\theta^{\text{false}}_{23})| >$ 2.71 and 6.63 
where $\theta^{\text{true}}_{23}$ and 
$\theta^{\text{false}}_{23}$ are the true and the false values of 
$\theta_{23}$, respectively. 
The solid (dashed) curves in Fig.~\ref{region_resolved1} and 
Fig.~\ref{region_resolved2} are for cases that we fit the data set 
generated with $\Delta m^2_{31}>0$ under the hypothesis of 
right hierarchy, $\Delta m^2_{31}>0$ (wrong hierarchy, 
$\Delta m^2_{31}<0$). 
If we are ignorant about the mass hierarchy the case of worse sensitivity  
(wrong hierarchy) must be considered as the sensitivity region.  
The fact that the solid and the dashed curves come close to each 
others proves  that the resolution of the $\theta_{23}$ octant degeneracy 
can be carried out independent of the lack of knowledge on the 
neutrino mass hierarchy, the decoupling of the two degeneracies 
as argued in Sec.~\ref{decoupling}.

The figures indicate that our method of combining reactor measurement 
of $\theta_{13}$ with the accelerator disappearance and appearance 
experiments allows to resolve the $\theta_{23}$ octant degeneracy 
to a reasonable level. 
This is highly nontrivial because it is quite a robust degeneracy which is 
hard to lift by using only relatively short-baseline ($\sim 1000$ km) 
accelerator measurement, as we saw in Sec.~\ref{robust}.
We can observe by comparison between 
Fig. \ref{region_resolved1} and Fig. \ref{region_resolved2} 
that improvement of the systematic errors in reactor measurement 
is the key to the better resolving power of the octant degeneracy. 
It should also be mentioned that the results in
Fig. \ref{region_resolved2} are obtained with the extremely small
systematic errors.  Hence, the results may be interpreted as the limit
achievable by the present method.

So far we have assumed that the input mass hierarchy is normal 
while examining the true and the false mass hierarchies 
in analyzing data. 
In order to verify that resolving power of $\theta_{23}$ degeneracy 
is not affected by the unknown sign of $\Delta m^2_{31}$, 
we tried the converse. Namely, 
we tried to fit the data set generated with the inverted hierarchy 
$\Delta m^2_{31}<0$ under the hypothesis of the right and the 
wrong hierarchies. 
We do not show the results because they are very similar to those 
for the input normal mass hierarchy.  
Of course, this is expected because the experimental setting we 
consider in this work is not sensitive to the sign of $\Delta m^2_{31}$, 
which in turn implies that our results must be insensitive to the mass 
hierarchy confusion. 
This point was discussed in depth in Sec.~\ref{robust}.

We note that globally our results in Fig.~\ref{region_resolved1} and 
Fig.~\ref{region_resolved2} are similar to what we can find in 
Fig.7 of Ref.~\cite{MSYIS}, which was obtained by a much more
simplified analysis.  
However, there are some notable differences. 
In Fig.7 of Ref.~\cite{MSYIS}, it is always true that the 
$\theta_{23}$ degeneracy is easier to lift in the second octant, 
$\theta_{23} > \pi/4 $. 
But, we observe that it is true only for relatively small $\theta_{13}$, 
$\sin^2 2\theta_{13} \lsim 0.06$. 
For relatively large $\theta_{13}$, $\sin^2 2\theta_{13} \gsim 0.06$, 
the octant degeneracy is easier to be resolved for the first octant 
case $\theta_{23} < \pi/4$ than for the second octant. 

It appears that at large $\theta_{13}$, the two effects which were not 
taken into account in the treatment in \cite{MSYIS} come into play, 
as indicated in Fig.~\ref{23fig1}; 
The width of the appearance strip, which comes from $\delta$-dependent 
terms in the oscillation probability (see Eq.~(\ref{Pmue}) in Sec.~\ref{analytic}), 
is narrower at smaller $\theta_{13}^{\rm fake}$, 
which is the case of true $\theta_{23}$ in the first octant, 
making rejection of the fake solution easier.  
Also, the two disappearance ``lines'' come closer in the case of true 
$\theta_{23}$ in the second octant, which produces the similar effect. 
As a result of the two effects which simultaneously act toward the same 
direction, the degeneracy turns out to be easier to resolve for the 
case of true $\theta_{23}$ in the first octant. 
We, however, want to note that the feature of first-second octants 
asymmetry, in particular in Fig.~\ref{region_resolved1}, depends upon 
the systematic errors and its full understanding may require much 
more subtle discussions.

\section{Concluding remarks}\label{conclusion}

In this paper, we have explored the possibility of resolving 
the $\theta_{23}$ octant degeneracy by combining reactor 
measurement of $\theta_{13}$ with the possible highest accuracy 
accelerator disappearance and appearance measurement, 
as proposed in \cite{MSYIS}.  
It utilizes the nature of the reactor experiment as a pure 
measurement of $\theta_{13}$ to resolve the degeneracy. 
It is nice to see that the reactor measurement can contribute 
to explore the two small quantities in lepton flavor mixing, 
$\theta_{13}$ and a deviation of $\theta_{23}$ from the maximal value.
The results of our quantitative analysis indicates reasonably 
high performance of the method. 
As shown in the summary figures of the resolving capabilities,  
Fig.~\ref{region_resolved1} and Fig.~\ref{region_resolved2}, 
the degeneracy is resolved in region of relatively large 
$\theta_{13}$ and a sizable deviation of $\theta_{23}$ from the maximal.

Prior to the quantitative analysis based on our method, 
we have discussed the robustness 
of the $\theta_{23}$ octant degeneracy and illuminated 
the difficulty in resolving it only by accelerator experiments. 
In particular, we have demonstrated by showing a sample figure, 
Fig.~\ref{app-disapp}, that neither the spectral information 
nor the matter effect enables us to resolve the degeneracy, 
This feature was also understood on the basis of analytic treatment 
using the approximate formulas for appearance and disappearance 
probabilities valid to first order in matter effect. 
Considering the robustness of the degeneracy, the opportunity 
of resolution offered by our reactor-accelerator method is highly 
nontrivial.

Finally, several remarks are in order:

\vspace{0.3cm}
\noindent
(1) As indicated in Fig.~\ref{23fig1}, 
the sensitivity of our method for resolving the $\theta_{23}$ 
degeneracy is limited mainly by the accuracy of $\theta_{13}$ 
determination by reactor experiments. 
The region without resolving power which remains in 
Fig.~\ref{region_resolved2} must be taken as the intrinsic 
limitation of our method for 
resolving the $\theta_{23}$ degeneracy in its current form, 
because we have already assumed a rather optimistic values of 
systematic errors in the reactor measurement,  
in addition to the extreme accuracies of accelerator experiments.

\vspace{0.3cm}
\noindent
(2) In this paper, we have considered the setting of T2K experiment 
originally described in \cite{JPARC}. 
It would be interesting to examine how the sensitivity of 
resolution of $\theta_{23}$ degeneracy changes if we adopt the 
Kamioka-Korea identical two-detector setting by which the 
degeneracy related to the neutrino mass hierarchy and the 
intrinsic one can be resolved \cite{T2KK}.

\vspace{0.3cm}
\noindent
(3) Our method for resolving $\theta_{23}$ degeneracy is 
by no means unique. 
The other possibilities include: 
accelerator measurement with silver channel 
($\nu_{e} \rightarrow \nu_{\tau}$) \cite{silver} 
which has a different $\theta_{23}$ dependence, and 
detection of the solar oscillation term by either  
atmospheric neutrino observation, 
\cite{atm23,concha-smi_23,choubey2}
or 
very long baseline accelerator experiments \cite{BNL}. 
Combination of the accelerator and the atmospheric neutrino 
experiments can also be pursuit \cite{atm-lbl}.
The ultimate possibility would, of course, be the ``everything at once'' 
approach \cite{donini} which combines measurement by 
superbeam and neutrino factories with golden and silver channels.

\appendix

\section{Calculation of Number of Events}\label{number}

In this appendix, we provide some detailed informations about 
how the number of events are computed. 
For clarity of notation we denote the neutrino energy ($E$ in the text) 
as $E_{\nu}$ in this appendix.

\vglue 0.2cm
\noindent
(i) {\bf Apparence channel $\nu_\mu \to \nu_e$ 
(or $\bar{\nu}_\mu \to \bar{\nu}_e$)}
\vglue 0.5cm
As we mentioned in sec. IV, since the information on the energy spectrum 
of the appearance channel is not important in resolving $\theta_{23}$ 
degeneracy, for this channel, we do not try to make binning but 
use the expected total number of events, which is computed as,  
\begin{equation}
N_{\rm sig}  =  
n_N T \int_{E_\nu^{\rm min}}^{E_\nu^{\text{max}}}
dE_\nu \phi_{\nu_\mu}(E_\nu) 
P(\nu_\mu\to \nu_e;E_\nu)
\sigma^{\nu_e}_{\rm tot}(E_\nu) 
\epsilon_{\nu_e}(E_\nu),
\end{equation}
where $E_\nu$ is the (true) neutrino energy, 
$n_N$ is the number of target nucleons in the detector,  
$T$ is the running (exposure) time,  
$\phi_{\nu_\mu}(E_\nu)$ is the flux spectrum of
the 2.5 degree off axis $\nu_\mu$  beam 
$\sigma^{\nu_e}_{\rm tot}(E_\nu)$ is the
total cross section and  
$\epsilon_{\nu_e}(E_\nu)$ is 
the detection efficiency, which is  
given as a function of the neutrino energy. 
The detection efficiency we used takes into account all 
the cut imposed to reduce background, which enables us to 
obtain the signal distribution in the range between 
0.35 and 0.85 GeV in terms of reconstructed neutrino 
energy \cite{JPARC-detail}.

We consider two kinds of background events, 
\begin{equation}
N_{\rm total}^{\rm BG} 
 =  N_{\rm total}^{\rm BG(NC)} + N_{\rm total}^{\rm BG(beam)}, 
\end{equation}
where $N_{\rm total}^{\rm BG(NC)}$ comes from neutral current 
interactions and 
$N_{\rm total}^{\rm BG(beam)}$ implies the event induced by 
the $\nu_e$ neutrino existed in the original beam. 
As for the signal, we used the efficiency functions which correspond 
to the cuts found in Ref.~\cite{JPARC-detail} in order to compute 
the background.  
\vglue 0.5cm 

\vglue 0.5cm 
\noindent
(ii) {\bf Disapparence channel $\nu_\mu \to \nu_\mu$ 
(or $\bar{\nu}_\mu \to \bar{\nu}_\mu$)}
\vglue 0.2cm 
For the disappearance mode, it is essential to make binning in order to 
take into account the information on the energy spectrum. 
The expected number of signal events for the i-th bin 
is computed as follows 
\begin{equation}
N_{\rm sig, i}  =  
n_N T \int_{E_{\rm rec}^{\rm i; min}}^{E_{\rm rec}^{\text{i; max}}}
dE_{\rm rec}
\int_{E_\nu^\text{min}}^{E_\nu^\text{max}} dE_{\nu}
 \phi_{\nu_\mu}(E_\nu) 
P(\nu_\mu\to \nu_\mu; E_\nu)
\sigma^{\nu_\mu}_{\rm CCQE}(E_\nu) 
\epsilon^{\rm rec}_{\nu_\mu}(E_{\rm rec})
R(E_\nu,E_{\rm rec}),
\end{equation}
where $E_{\rm rec}$ implies reconstructed neutrino energy, 
$\sigma^{\nu_\mu}_{\rm CCQE}(E_{\rm rec})$ 
implies charged current quasi elastic (CCQE) 
reaction cross section, 
$\epsilon^{\rm rec}_{\nu_\mu}(E_{\rm rec})$ is the 
detection efficiency as a function 
of the reconstructed neutrino energy, which corresponds
to the one used in Ref.~\cite{Hiraide},  
$R(E_\nu,E_{\rm rec})$ is the Gaussian-like 
resolution function, 
\begin{equation}
R(E_\nu,E_{\rm rec}) = \frac{1}{\sqrt{2\pi} \sigma_E}
\text{exp} 
\left[ 
-\frac{1}{2}\left(\frac{E_\nu-E_{\rm rec}}{\sigma_E}\right)^2 
\right], 
\end{equation}
where we set $\sigma_E$ = 80 MeV.

For this mode, we consider two kinds of background, 
\begin{equation}
  N_{i}^{\rm BG} 
 =  N_{i}^{\rm BG(NC)} + N_{i}^{\rm BG(CC-NQE)}, 
\end{equation}
where $N_{i}^{\rm BG(NC)}$ is the events coming from neutral current
reaction and $N_{i}^{\rm BG(CC-NQE)}$ is the one coming from 
the charged current non-quasi elastic (CC-NQE) reactions. 
We note that $N_{i}^{\rm BG(NC)}$ do not depend on oscillation parameters,
whereas $N_{i}^{\rm BG(CC-NQE)}$ depend on 
oscillation parameters 
in a non-trivial way. 
For the background coming from CC-NQE reactions, 
$E_{\rm rec} = E_{\rm true}$ is not a good 
approximation because events induced by higher energy neutrinos
mimic the signal induced by lower energy ones. 
However, we observe that it is a good approximation to take that
$E_{\rm rec} - E_{\rm true} = 300$ MeV or 
$N_{i}^{\rm BG(CC-NQE)}(E_{\rm rec}) \simeq 
N_{i}^{\rm BG(CC-NQE)}(300\ {\rm MeV}+ E_{\rm true})$. 
This allows us to compute 
$N_{i}^{\rm BG(CC-NQE)}(E_{\rm rec})$ in the presence of 
oscillation provided that we 
know the distribution $N_{i}^{\rm BG(CC-NQE)}$ 
as a function of reconstructed neutrino energy in the 
absence of oscillation for the 2.5 degree beam. 

\vglue 0.5cm 
\noindent
(iii) {\bf Disappearance channel $\bar{\nu}_e \to \bar{\nu}_e$}
\vglue 0.5cm 
We compute the expected number of $\bar \nu_e$ events in the 
$i$-th energy bin, $N_i^{\text{theo}}= N_i^{\text{reac}}$, 
where $N_i^{\text{reac}}$ 
\begin{eqnarray}
N_i^{\text{reac}}(\sin^2 \theta_{12},\Delta m^2_{21}) &=&
N_p T 
\nonumber \\
\times \int &dE_\nu& \, \epsilon
\phi(E_\nu) P(\bar{\nu}_e \to \bar{\nu}_e; L, E_\nu) 
\sigma(E_\nu) \int_i dE 
\epsilon_{\text{det}}\,R(E,E^\prime),
\end{eqnarray} 
where $N_p$ is the number of target protons in the detector 
fiducial volume,  $T$ is the exposure time, and 
$\phi(E_\nu)$  is the neutrino flux spectrum from 
the nuclear power plant (NPP) expected at its maximal thermal 
power operation.
$\epsilon$ denotes the averaged operation efficiency of 
the NPP for a given exposure period and it is 
taken to be 100\% here under the understanding that the unit we use 
GW$_{\text{th}}\cdot$kt$\cdot$yr refers the actual thermal power 
generated, not the maximal value. 
$P(\bar{\nu}_e\ \to \bar{\nu}_e, L, E_\nu)$ is the familiar 
antineutrino survival probability in vacuum, 
which is given by Eq.(\ref{Pvac_ee})
and it explicitly depends on $\Delta m^2_{21}$ and  $\sin^2 \theta_{12}$.
$\sigma(E_\nu)$ is the $\bar{\nu}_e$ absorption cross-section 
on proton, $\epsilon_{\text{det}}=0.898$ is the detector efficiency and 
$R(E,E')$ is the energy resolution 
function, which is assumed to have a Gaussian form with 
$E$ ($=E_{\text{prompt}})$ the observed prompt energy (total $e^+$ energy) 
and $E'=E_{\nu}$ - 0.8 MeV 
the true one.

\begin{acknowledgments}
Two of us (H.M. and H.N.) thank Takaaki Kajita with whom our
understanding on parameter degeneracies has been deepened through the
collaboration for \cite{T2KK}.
This work was supported in part by the Grant-in-Aid for Scientific Research, 
No. 16340078, Japan Society for the Promotion of Science (JSPS), 
Funda\c{c}\~ao de Amparo \`a Pesquisa do  Estado de S\~ao Paulo 
(FAPESP),  
Funda\c{c}\~ao de Amparo \`a Pesquisa do  Estado de Rio de Janeiro 
(FAPERJ) and Conselho Nacional Nacional de Ci\^encia e Tecnologia (CNPq). 
H.S. thanks JSPS for support.
R.Z.F. is grateful to the Abdus Salam International Center for
Theoretical Physics where part of this work was done.
\end{acknowledgments}

\end{document}